\documentclass[10pt,letterpaper]{article}
\usepackage[top=0.85in,left=2.75in,footskip=0.75in]{geometry}

\usepackage{amsmath,amssymb}

\usepackage{changepage}

\usepackage[utf8x]{inputenc}

\usepackage{textcomp,marvosym}

\usepackage{cite}

\usepackage{nameref,hyperref}

\usepackage[right]{lineno}

\usepackage{microtype}
\DisableLigatures[f]{encoding = *, family = * }

\usepackage[table]{xcolor}

\usepackage{array}

\newcolumntype{+}{!{\vrule width 2pt}}

\newlength\savedwidth

\newcommand\thickhline{\noalign{\global\savedwidth\arrayrulewidth\global\arrayrulewidth 2pt}%
\hline
\noalign{\global\arrayrulewidth\savedwidth}}

\usepackage{ragged2e}
\usepackage{fixltx2e}
\raggedright
\usepackage[aboveskip=1pt,labelfont=bf,labelsep=period,justification=raggedright,singlelinecheck=off]{caption}

\makeatletter
\renewcommand{\@biblabel}[1]{\quad#1.}
\makeatother

\date{}

\usepackage{lastpage,fancyhdr,graphicx}
\usepackage{epstopdf}
\fancyheadoffset[L]{2.25in}
\fancyfootoffset[L]{2.25in}
\newcommand{\lorem}{{\bf LOREM}}
\newcommand{\ipsum}{{\bf IPSUM}}

\newcommand{\ignore}[1]{}
\usepackage{color, colortbl}
	
\definecolor{Gray}{gray}{0.9}
\definecolor{LightCyan}{rgb}{0.88,1,1}

\begin{document}
\vspace*{0.2in}

\begin{flushleft}
{\Large
\textbf\newline{\texttt{FRETtranslator}: translating FRET traces into RNA structural pathways} 
}
\newline
\\
Nikolai Hecker\textsuperscript{1, 2},
Matthew Kahlscheuer\textsuperscript{3},
Peter Kerpedjiev\textsuperscript{5},
Jan Gorodkin\textsuperscript{1, 2},
Peter F. Stadler \textsuperscript{1, 5-10},
Ivo L. Hofacker\textsuperscript{1, 5, 6},
Nils Walter\textsuperscript{3*},
Jing Qin\textsuperscript{1, 5, 11*}
\\
\bigskip
\textbf{1} Center for non-coding RNA in Technology and Health\\
\textbf{2} Department of Veterinary Clinical and Animal Sciences, University of Copenhagen, Gr{\o}nneg{\aa}rdsvej 3, 1870 Frederiksberg C, Denmark\\
\textbf{3} Single Molecule Analysis Group, Center for RNA Biomedicine, Department of Chemistry, University of Michigan, Ann Arbor, USA \\
\textbf{4} Centre for Bioinformatics, University of Hamburg, Bundesstr. 43, 20146 Hamburg\\
\textbf{5} Institute for Theoretical Chemistry, Univ.\ Vienna, W\"{a}hringerstr.\ 17, 1090 Vienna, Austria\\
\textbf{6} Research group BCB, Faculty of Computer Science, Univ.\ Vienna, Austria\\
\textbf{7} Dept.\ of Computer Science \& IZBI \& iDiv \& LIFE, Leipzig Univ., 
        H{\"a}rtelstr.\ 16-18,  Leipzig, Germany\\
\textbf{8} Max Planck Institute for Mathematics in the Sciences,
        Inselstr. 22, Leipzig, Germany\\
\textbf{9} Fraunhofer Institute IZI, Perlickstr.\ 1, Leipzig, Germany\\
\textbf{10} Santa Fe Institute, 1399 Hyde Park Rd., Santa Fe, NM87501, USA\\
\textbf{11} IMADA, Univ.\ Southern Denmark, Campusvej 55, Odense, Denmark.
\\
\bigskip

%
%





* qin@tbi.univie.ac.at (JQ); nwalter@umich.edu (NGW) 
\end{flushleft}
\section*{Abstract}
\justify
Recent genome and transcriptome sequencing projects have unveiled a plethora of highly structured RNA molecules 
as central mediators of cellular function. Single-molecule F{\"o}rster Resonance Energy Transfer (smFRET) is a powerful 
tool for analyzing the temporal evolution of the full structure of individual RNA molecules, in pursuit of understanding 
their essential structure-dynamics-function relationships. In contrast to enzymatic and chemical footprinting, NMR 
spectroscopy and X-ray crystallography, smFRET yields temporally resolved, quantitative information about single 
molecules rather than only time and ensemble averages of entire populations. This makes unique observations of 
transient and rare conformations under both equilibrium and non-equilibrium conditions possible. However, the 
structural information obtained is limited to a one-dimensional (1D) distance between pairs of fluorophore labeling 
sites as a projection of a molecule's 3D structure. It therefore remains an unmet challenge to relate smFRET data 
back to the underlying temporal sequence of transitions between RNA 3D structures. Here, we describe a hidden 
Markov model approach to translate smFRET time traces into the most probable corresponding secondary structure 
paths on the folding free energy landscape of the RNA. This is achieved by combining predictions from the RNA's 
folding landscape and 3D structure into a computational framework termed \texttt{FRETtranslator}. \texttt{FRETtranslator} predicts 
RNA secondary structure transitions solely based on the RNA sequence and an input smFRET trace. Application of 
\texttt{FRETtranslator} to a preQ\textsubscript{1}-riboswitch successfully recapitulates its reversible transitions between hairpin and 
pseudoknot structures. 
\ignore{We then analyze the conformational dynamics of a 135-nucleotide pre-messenger RNA and cross-reference our predictions 
with independently derived footprinting data.} 
To our knowledge, \texttt{FRETtranslator} is the first computational tool that provides a direct structural interpretation of smFRET data for RNA. 
It is freely available at http://sourceforge.net/projects/\texttt{FRETtranslator}/ under terms of the GNU General Public License v3.0.

\section*{Author Summary}
Understanding how an RNA sequence folds is becoming increasingly important as our understanding of the diverse functions of RNA is continuously expanding. 
Measurements using single-molecule F{\"o}rster resonance energy transfer (smFRET) have yielded unprecedented kinetic and mechanistic insights into RNA folding 
and conformational changes. However, observations from smFRET experiments are limited to a time series of inter-fluorophore distances between two specific, 
pre-determined labeling sites that reflect but also hide many details of the underlying RNA structures and folding pathways. Here, we present \texttt{FRETtranslator}, 
a hidden Markov model approach that predicts the most likely time series of RNA secondary structure transitions underlying a given smFRET trace.  \texttt{FRETtranslator} 
combines the smFRET-derived inter-fluorophore distance information with coarse-grained modeling of both the folding free energy landscape and resulting 3D structures 
of the RNA. We applied \texttt{FRETtranslator} to smFRET data for a 36-nt
transcriptional  preQ\textsubscript{1} riboswitch from \textit{Bacillus
  subtilis} (\textit{Bsu})
to predict the temporal sequences of secondary structures underlying their equilibrium folding transitions. \texttt{FRETtranslator} is freely available at \\
http://sourceforge.net/projects/\texttt{FRETtranslator}/.


\section*{Introduction}
Key cellular processes involving the maintenance, modification, and regulated expression of the genome critically depend on the ability of RNA to fold into specific 
structures that are sufficiently dynamic to undergo rearrangements that are
necessary for their function. For instance, bacterial riboswitches change their conformation 
upon ligand binding, leading to an altered level of transcription or translation \cite{RN1,RN2,RN3,RN4}; RNA thermometers harness structural rearrangements to affect 
temperature-induced changes in gene expression \cite{RN5}; catalytic RNAs utilize magnesium-dependent changes in structure to gate catalysis \cite{RN6, RN7}; and the efficient 
biosynthesis of viable messenger (m)RNAs relies on guided conformational rearrangements of the pre-mRNA substrate of the spliceosome \cite{RN8, RN9, RN10}. 

Since the functions of these non-coding RNAs intrinsically tied to such structural rearrangements, it is important to determine and further 
investigate these structural changes and their folding kinetics. Extensive experimental and computational studies on RNA folding have provided 
significant insights into the kinetic mechanism of RNA functions, see \cite{RN11} and references therein.

Experimental approaches such as RNA footprinting or spectroscopy \cite{RN12, RN13}, have been developed to incorporate structural constraints in order to improve the accuracy 
for predicting RNA native structures. Both base-paired and unpaired nucleotides can be more accurately defined \cite{RN14, RN15}.  However, due to their inherent ensemble- and 
time-averaging, both footprinting and NMR spectroscopy typically obtain ambiguous or ill-defined data for dynamic RNAs, which cannot easily be deconvolved into the constituent 
time series of individual RNA structures.

Single-molecule fluorescence resonance energy transfer (smFRET) \cite{RN16} microscopy has become an increasingly popular tool to study the structural dynamics of RNA 
molecules. It reveals, in real time, the structural dynamics of these molecules  by monitoring the distance between two dye-labeled atoms of an RNA molecule. Unlike ensemble 
structure probing, smFRET reveals events occurring transiently and in small numbers of individual molecules both at and far from equilibrium. It provides information about 
subpopulations with distinct behaviors and can detect even subtle changes in folding kinetics that alter RNA tertiary or secondary structure. Hence, smFRET has been successfully 
used to characterize the role of dynamics in functional RNAs and their protein complexes, including riboswitches, ribozymes, the ribosome, and the spliceosome \cite{RN1, RN2, RN6, RN9, RN17, RN18, RN19}. 

Previous computational analyses of smFRET time traces have typically focused on characterizing the idealized FRET states and estimating the transition rates between them 
\cite{RN20, RN21}. Associating specific structures with these states, however, remains a challenging task and often requires arbitrary choices and/or expensive and time-consuming 
control  experiments. For example, a sophisticated hidden Markov model (HMM) based approach was used to study smFRET traces of a Diels-Alderase ribozyme \cite{RN22}. The derived 
transition states were assigned to separately predicted RNA secondary structures. Consequently, the association of FRET states with secondary structures was not an intrinsic part 
of the model, but based on the authors' interpretation. A separate set of computational tools was developed for the simulation of tertiary structures. For instance, the 36-nucleotide 
(nt) transcriptional \textit{Bacillus subtilis} (\textit{Bsu}) riboswitch with a known crystal structure was studied with two coarse-grained molecular dynamics simulation approaches, termed 
TOPRNA and G\={o} models \cite{RN19, RN23}. In the former approach, predicted mean distances between the fluorophores were compared to measured smFRET efficiencies to assign a 
most likely structural model to a FRET state. However, molecular simulations of RNA usually require extensive a priori knowledge of the structure under investigation, e.g., a crystal 
structure \cite{RN24}. Without such prior knowledge, molecular simulations rely on an accurate prediction of the tertiary structure, which is notoriously unreliable in the absence of 
experimental input, given the rugged folding landscapes of RNAs \cite{RN25, RN26}. A recently developed method, multidimensional Chemical Mapping (MCM) \cite{RN27} provides a means of disentangling 
structural alternatives but lacks temporal information. At present MCM appears to be limited to abundant states, however.

Compared to these approaches, \texttt{FRETtranslator}'s advantage is that it predicts RNA secondary structure transitions  
based on \textit{solely} the RNA sequence and smFRET traces.  \texttt{FRETtranslator} employs an HMM approach to ''translate''  each input smFRET trace inindependently to a time series of specific 
RNA conformational changes. The state space for this HMM is a particular set of secondary structures in a coarse-grained model of the RNA folding landscape known as the basin 
hopping graph (BHG) \cite{RN28}. The transition probabilities between two states are evaluated computationally according to their corresponding energy barrier within the BHG framework 
as well. The emission states are grouped FRET values extracted from input smFRET traces. In the current study, input smFRET traces are idealized with well-established HMM software 
such as \texttt{vbFRET} \cite{RN20} and \texttt{QuB} \cite{RN21}.  The emission probability of a given secondary structure is inferred from an empirical distribution of inter-fluorophore distances 
by applying the F\"{o}rster equation. This empirical inter-fluorophore distance distribution is simulated with software \texttt{Ernwin} \cite{RN29}, which is able to efficiently sample 3D 
structures satisfying an input secondary structure. 

In the following, we refer to the resulting HMM as \textit{BHG-HMM} to distinguish it from the HMMs used in idealizing the smFRET input 
traces. To validate our method, we show that our predicted structural transitions for the \textit{Bsu} riboswitch example are consistent with those from previous TOPRNA modeling of 
similar smFRET data \cite{RN19}. 

\section*{Materials and Methods}
\subsection*{BHG-HMM Construction}
\subsubsection*{Hidden states, initial and transition probabilities from RNA folding landscapes}
We first briefly review the relevant concepts of the RNA folding landscape and the BHG approach. For details, we refer to \cite{RN28, RN30, RN31} and  \nameref{S1_Text}. 

Given an RNA sequence, we consider the underlying structural space as the set of all secondary structures that can be formed by this sequence assuming that 1) only canonical 
(\textsf{GC},\textsf{AU}, and \textsf{GU}) base pairs are formed; 2) hairpin loops have a minimum length of three; and 3) particular types of pseudoknots defined in \cite{RN32} can be included. These 
conditions define the ensemble of structures as implemented in the most commonly used RNA folding tools such as \texttt{Mfold} \cite{RN33} and the \texttt{ViennaRNA} package \cite{RN34}.

The (free) energy $f(x)$ of each structure $x$ in this ensemble is evaluated following well-established energy models \cite{RN35, RN36} in the form of additive contributions for 
base pair stacking as well as hairpin loops, interior loops, bulges and multi-loops. RNA structures in the underlying structural space are arranged as a graph by specifying which 
pair of structures can be interconverted with elementary rearrangements, typically the opening or closing of individual base pairs. This graph is referred to as the \textit{landscape} in the following. 

The size of the underlying space grows exponentially according to the RNA
sequence length \cite{RN37}. The coarse-grained basin hopping graph (BHG)
model was introduced in \cite{RN28} to alleviate this combinatorial
explosion and allow an efficient investigation of the folding dynamic of
RNAs of moderate lengths. Nodes in the BHG represent secondary structures
that are locally stable. More precisely, the folding energy of a structure
that qualifies as a BHG node is strictly smaller than the energies of all
neighbor structures in the landscape. We refer to such a structure as a
\textit{local minimum (LM)}. Two LMs in the BHG are linked by an edge if the direct
transition between them is ''energetically favourable", i.e., introducing
any detour in the landscape between these two LMs does not further lower
the energy required for the molecule to complete the conformational
conversion. The weight on each edge indicates the energy barrier for
converting two adjacent LMs. As a result, optimal folding pathways are
modeled as paths in the BHG represented by their LMs.

\texttt{FRETtranslator} uses the resulting BHG to obtain the hidden states, initial and transition probabilities. The hidden states are the LMs in the BHG, which we refer to as \textit{candidate 
structures}. The initial probabilities for the candidate structures are estimated from their Boltzmann weights. To be precise, for a candidate structure $x$ with free energy $f(x)$, its initial 
probability is 
$$\frac{e^{-f(x)/RT}}{\sum_{s\in \mathcal{X}}e^{-f(s)/RT} },$$
where the denominator sums over the set $\mathcal{X}$ of all candidate structures taking into consideration. 
We further describe the dynamics among candidate structures as a continuous-time Markov process. Assume there are in total $N$ candidate structures, transition probabilities 
between pairs of candidate structures for each time period t are then computed by numerically computing the matrix exponential $\exp{(t\mathbf{M})}$ where $\mathbf{M}=(r_{yx})$ 
is the $N\times N$ 
infinitesimal generator matrix \cite{RN30}. In which, $r_{yx}$ denotes the transition rate from a candidate structure $x$ to another candidate structure $y$. This rate is nonzero only if x and y are connected by exactly one edge in the BHG. In this case, according to the Metropolis rule,
\begin{equation}\label{E:tranprob}
r_{yx}= c_0 \cdot e^{-(S(x,y)-f(x))}
\end{equation}
$S(x,y)$ is the edge weight between $x$ and $y$ in BHG, $f(x)$ evaluates the energy of $x$, $R$ is the universal gas constant, $\mathcal{T}$ is the absolute ambient temperature. The 
parameter $c_0$ gauges the time axis and will be referred to as \textit{time-scalar} in the following. It collects entropic terms that cannot be computed directly from the landscape in our 
setting and depends to a certain extent on the particular given RNA sequence.  We treat this time scalar $c_0$ as a user-defined parameter. In our experience, a default value of 100 
is useful as a first trial when the unit time is 0.1s (see Result section). Such default value assumes that in 0.1s the molecule experiences about 100 basin-hoppings in its 
landscape.

\subsubsection*{Emission states and estimations of emission probabilities based on 3D RNA sampling}

The FRET (efficiency) values will be used as the emission states of our BHG-HMM. To reduce the influence of noise from the smFRET measurements, we use idealized smFRET 
traces as input, i.e., traces that are pre-processed with software such as vbFRET \cite{RN20} or QuB \cite{RN21}. These programs allow for the application of HMM algorithms to 
smFRET trajectories, and the extraction of idealized FRET states and rate constants of their interconversion. Also, these HMM programs currently present the most accessible form 
of data analysis that produces the most reliable results with minimal a priori assumptions required from the user \cite{RN38}. 

More precisely, with a user-defined parameter $b$, all nonzero FRET values are partitioned into b emission states: $(0, 1/b], \dots, (1-1/b, 1]$. In the following, we refer to these 
emission states as bins. For instance, $b=5$ indicates the FRET values grouped into five emission states $(0, 0.2]$, $(0.2, 0.4]$, $(0.4, 0.6]$, $(0.6, 0.8]$ and $(0.8, 1]$. We also use an additional bin to accommodate extreme situations like ``no signal'' or missing data. 

\begin{figure}[!h]
\centering
  \includegraphics[width=\textwidth]{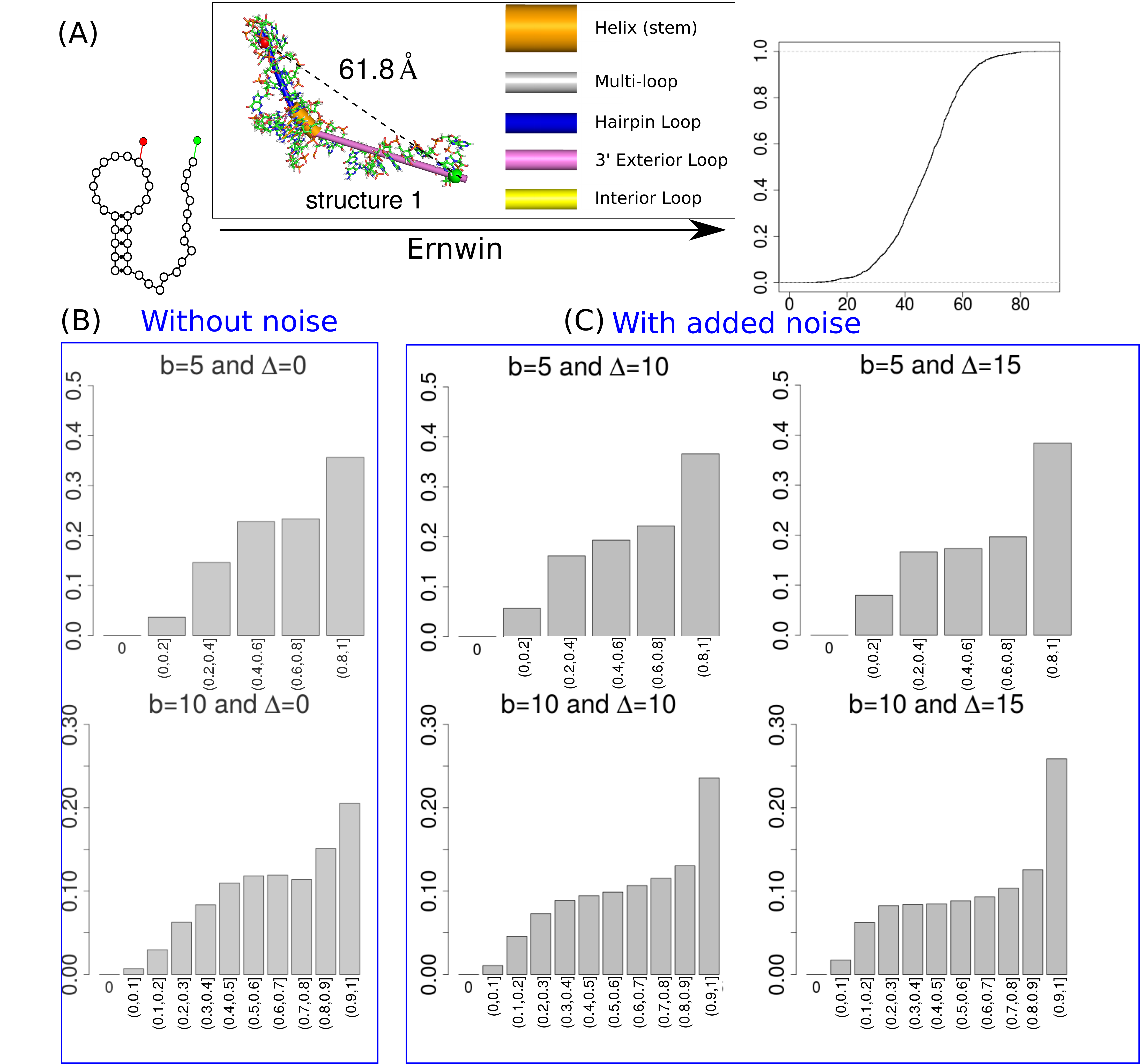}
\caption{{\bf Estimating emission probabilities of a candidate structure.} 
(A) Once a candidate structure is found, \texttt{Ernwin} \cite{RN29} is used to generate a set of \textit{in silico} 3D structures satisfying the structural constraints implied by this candidate structure. 
An examplary 3D structure is shown above the arrow with structural elements indicated in the legend. Afterwards, the inter-fluorophore distance distribution is collected from this set 
of \textit{in silico} 3D structures. The empirical cumulative distribution of the example candidate structure is shown to the right of the arrow. This inter-fluorophore distance distribution is 
then converted to a distribution of FRET values according to the F\"{o}rster equation. (B) Finally, grouping FRET values into b bins derives emission probabilities as shown in the 
histograms. Each bar indicates the emission probability of a bin and is labeled by the interval of FRET values that the corresponding bin contains.  (C) This derived distribution of 
emission probabilities can be perturbed by adding noise at a defined level as described in more detail at the end of this section. The number of bins and the noise level are user-
defined parameters. In this figure, we set the number of bins b to either 5 or 10 and the noise level $\Delta$  to 10 or 15 \AA. }
\label{F:estemissionprob}
\end{figure}

To estimate the emission probabilities, as exemplified by Fig.~\ref{F:estemissionprob} (A), \texttt{FRETtranslator} utilizes the following procedure. First, for each candidate structure, a set of 3D structures 
that satisfies the constraints implied by the candidate secondary structure is generated using \texttt{Ernwin} \cite{RN29}. We will refer to these sampled 3D structures as \textit{in silico 3D structures}. 
Next, we estimate \textit{in silico}  the distance $d$ between the locations where two fluorophores are located in each \textit{in silico} 3D structure. This yields the inter-fluorophore distance 
distribution for each candidate structure, which is further converted into the distribution of FRET values $E_{FRET}$ of the candidate structure according to the following F\"{o}rster 
equation \cite{RN39}: 
\begin{equation}\label{E:foerstereqn}
E_{FRET}=((1+(d/R_0)^6))^{-1}.
\end{equation}
Here, $R_0$ is the F\"{o}rster radius, i.e., the distance between the
fluorophores corresponding to 50\% FRET efficiency.  In \texttt{FRETtranslator}, we
use a default value of  $R_0=54$ \AA when the cyanine dyes Cy3 and Cy5 fluorophores are used \cite{RN19, RN40}. 

As mentioned above, all \textit{in silico} 3D structures are sampled with \texttt{Ernwin}, a coarse-grain helix-centered model that allows one to rapidly sample a set of 3D structures satisfying a given
secondary structure. Because of its coarse-graining nature, on one hand, \texttt{Ernwin} samples the conformational space more efficiently than approaches based on all-atom models 
such as the Rosetta software suite \cite{RN41}. Therefore, sufficient samplings of \textit{in silico} 3D structures can be derived for larger RNA molecules. On the other hand, inter-fluorophore 
distances measured by \texttt{Ernwin} are approximations of distances between coarse-grain structural elements where the fluorophores are located. For more details regarding this 
approximation, see \nameref{S2_Text}. 

Therefore, we assume that certain inaccuracies can exist in the \texttt{Ernwin}-estimated distribution of inter-fluorophore distances and the resulting emission probabilities. To address this 
inaccuracy, noise at different levels is added to ``perturb" both the derived distribution of inter-fluorophore distances and the structural transition results predicted with \texttt{FRETtranslator}. 
Only predictions robust against this added noise are taken into consideration.

More particularly, for each inter-fluorophore distance d obtained with \texttt{Ernwin}, 100 random noise values within a user defined range $(d-\Delta, d+\Delta)$ are added to the original 
\texttt{Ernwin} derived data. These random values are generated following by default a uniform distribution. In Fig.~\ref{F:estemissionprob} (B), we show an example of emission probabilities 
estimated without and with added noise. Noise generated by other distributions, for instance a Gaussian distribution, can be handled analogously. The user-defined range $\Delta$ 
is referred to as \textit{noise level} throughout the remainder of this contribution. When no noise is added, we simply write $\Delta=0$.

\subsection*{The Viterbi algorithm}
For a given RNA molecule, BHG-HMMs used in \texttt{FRETtranslator} can be constructed based on BHG and \texttt{Ernwin}-sampled distances as described above. The resulting model includes 
the following information: a set $S$ of candidate structures, initial probabilities $\pi(s_1)$  for starting from the candidate structure $s_1$, transition probabilities $\alpha(s_i, s_j)$ of 
transitioning from candidate structure $s_i$ to $s_j$, and emission probabilities $\beta(F_k\vert s_i )$ that one observes a FRET value within bin $F_k$ given a candidate structure 
$s_i$. In which,  $F_k$ is one of the $b$ bins mentioned above. 

Assume a BHG-HMM is fixed and an input FRET trace with FRET values of length $T$ (time steps), denoted by $f_1, f_2, \dots, f_T$, is given. \texttt{FRETtranslator} first converts each 
FRET value $f_t$ to its corresponding bin $F_t$, where $1\leq t\leq T$. Then the likelihood function for a sequence of candidate structures $s_1, s_2, \dots ,s_T$ that produces this 
FRET trace is computed in the following:
\begin{eqnarray*}
&&\mathbb{P}(s_1,s_2, \dots,s_T\vert F_1,F_2, \dots F_T )\\
&=&\pi(s_1 )\cdot \beta(F_1\vert s_1 )\cdot \alpha(s_1,s_2 )\cdot \beta(F_2\vert s_2 )\cdot \dots \cdot 
\beta(F_{T-1}, s_{T-1} )\cdot \alpha(s_{T-1},s_T )\cdot \beta(F_T\vert s_T).
\end{eqnarray*}
The most likely sequence of candidate structures $x_1,x_2, \dots x_T$  is thus the one that maximizes this likelihood. We refer to the logarithm of this maximal likelihood as the 
\textit{Viterbi score} of this input FRET trace under the specific BHG-HMM architecture. The Viterbi score of this BHG-HMM is defined as the sum of the Viterbi scores over all input 
traces.

The Viterbi algorithm \cite{Forney:73} derives the optimal sequence of candidate structures with time complexity $\text{O}(T\cdot \vert S\vert^2)$ in dynamic programming manner, 
where $T$ is the length of the FRET trace and $\vert S\vert$ is the total number of candidate structures taken into consideration. 

\subsection*{\texttt{FRETtranslator} implementation details}

\texttt{FRETtranslator} employs StochHMM \cite{RN42} for the Viterbi-decoding. \texttt{FRETtranslator} provides a collection of Perl, R and BASH scripts that are compliant with Linux and 
most UNIX-like systems for analyzing pre-processed smFRET traces.

\subsection*{Parameter selection and robustness analysis in the \textit{Bacillus subtilis} preQ\textsubscript{1} riboswitch case study}

As mentioned above, different variants of the BHG-HMMs can be constructed by choosing a combination of the three user-defined parameters: the number of bins, $b$, used to group the FRET values, the 
time-scalar, $c_0$, used to gauge the time axes in Eqn.~\ref{E:tranprob} and the noise level, $\Delta$, used to perturb distributions of \textit{in silico} distances and their resulting emission probabilities. A na{\"\i}ve 
method for model selection is to evaluate the quality of each BHG-HMM according to the sum of the Viterbi scores over all the input FRET traces and then to choose the model with the maximal value.

In principle, one could try as many combinations of the BHG-HMM parameters as possible, however, this is computationally demanding and not necessary since in practice many parameter combinations give 
rise to similar results. For this reason, we limited ourselves to a fixed number of sparse parameter combinations: We set the number of bins $b$ to be either 5 or 10, the noise level $\Delta$ to be 0, 10\AA 
or 15\AA, and used 4 values for the the time-scalar $c_0$ for the \textit{Bsu} preQ\textsubscript{1} riboswitch.
\ignore{either 4 (\textit{Bsu} preQ\textsubscript{1} riboswitch) or 5 (Ubc4 pre-mRNA) values for the time-scalar, $c_0$.} Further, it is not ideal to compare BHG-HMMs with different numbers of bins because, 
when there are more emission states, the emission probabilities get lower and so is the overall likelihood for a particular trace. 

\begin{figure}[!h]
\centering
  \includegraphics[width=\textwidth]{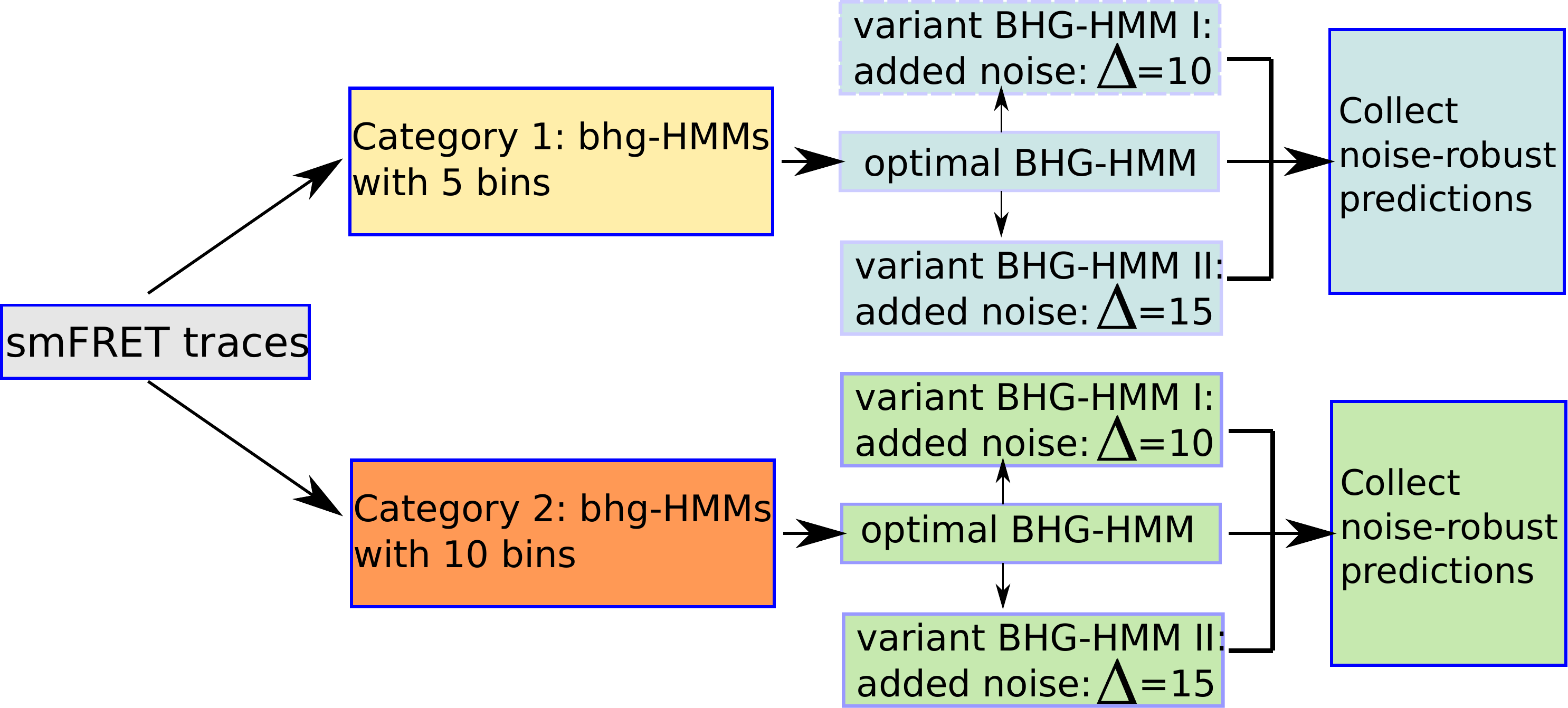}
\caption{{\bf Work flow in case study.} 
We analyzed our two examples according to a two-step strategy: First, without adding noise, we grouped the parameter combinations into two categories according to the number of bins. Then for each 
category, only the optimal BHG-HMM with the highest Viterbi score was selected for the second step featuring a more robust analysis. In this second step, input FRET traces were further analyzed with 
\texttt{FRETtranslator} using noise-perturbed variants of the original optimal BHG-HMMs to extract robust structural information. In this step, we considered two levels of added noise, 10 and 15 \AA. }
\label{F:workflow}
\end{figure}

Thus, we analyzed our two RNAs following the two-step strategy illustrated in Fig.~\ref{F:workflow}. First, without adding noise, we grouped the parameter combinations into the two categories according to the 
number of bins. Then for each category, only the optimal BHG-HMM with the highest Viterbi score was selected for the second step of a robustness analysis. In the second step, input FRET traces were further 
analyzed with \texttt{FRETtranslator} using noise-perturbed variants of the original optimal BHG-HMMs to extract robust structural information. In other words, we aim at identical predictions that disregard changes in 
noise levels. In this step, we considered two levels of added noise: $\Delta = 10$ and 15\AA.

\subsection*{Data preparation for the \textit{Bsu} preQ\textsubscript{1} riboswitch case study.}
\subsubsection*{Preparation of RNAs for smFRET.}
The 36-nt \textit{Bsu} RNA construct was synthesized by Dharmacon Inc. containing a 5-biotin modification for immobilization, a 3'DY547 fluorescent label, and a 5'aminoallyl-uridine (5-N-U) at position U13 for 
functionalization with Cy5 (Table \ref{T1:seq}) \cite{RN19}. Upon deprotection, 3.4 nmol RNA was incubated with one dye pack of Cy5-HNS ester (GE Healthcare) dissolved in 30 $\mu$l DMSO in a total 
reaction volume of 50 $\mu$l containing 0.1 M sodium bicarbonate buffer (pH 8.7). Reactions were allowed to proceed at room temperature for 4 h before dilution to 500 $\mu$l with water and purification on a 
Nap-5 gel filtration column (GE Healthcare). Fractions containing the labeled RNA were collected and ethanol precipitated. 

\begin{table}[!ht]
\centering
\caption{
{\bf Sequence information for the \textit{Bsu} preQ\textsubscript{1} riboswitch case study. \ignore{RNA constructs used in this study.}}}
\begin{tabular}{|l|}
\hline
5'-AGAGGUUCUAGC ({\color{red}{5-N-U}}) ACACCCUCUAUAAAAAACUAAGG ({\color{green} DY547})-3' \\ \hline
\end{tabular}
\begin{flushleft} 
\ignore{The allyl-amine modified uridines utilized for labeling are denoted as (5-N-U) with red and green colors representing positioning of the Cy5 and Cy3 fluorophores, respectively.} The \textit{Bsu} preQ\textsubscript{1} riboswitch was synthesized with the Dy547 fluorophore on its 3' end to serve as the donor fluorophore.
The allyl-amine modified uridines utilized for labeling is denoted as (5-N-U) with red representing positioning of the acceptor fluorophore. 
\end{flushleft}
\label{T1:seq}
\end{table}

\subsubsection*{smFRET analysis of RNA constructs.}
smFRET analysis of the \textit{Bsu} preQ\textsubscript{1} riboswitch RNA was performed as previously described in \cite{RN19}. Briefly, the RNA was folded by heating at 70 $^{\circ}\mathrm{C}$ for 2 min and allowing it to cool to room temperature (RT) for at least 20 min in smFRET buffer (50 mM Tris-HCl (pH 7.5), 100 mM KCl) in the presence of 100 nM preQ\textsubscript{1}. Folded RNA was flowed onto a quartz slide coated with biotinylated-BSA and streptavidin and allowed to incubate for 10 min for binding. Excess RNA was removed by washing with 200 $\mu$l of smFRET buffer with preQ\textsubscript{1}. smFRET was carried out using a prism-based TIRF microscope, a 532-nm laser to excite the donor (Cy3), and a 635-nm laser to excite the acceptor (Cy5) with the emission recorded at 100 ms time resolution with a Princeton Instruments, I-PentaMAX intensified CCD camera. Data were collected in the presence of the smFRET buffer containing an oxygen scavenging system (OSS) composed of protocatechuate dioxygenase, protocatechuate and Trolox by directly exciting Cy3 and recording of both Cy3 and Cy5 emission intensities. Following molecule selection, the $k$-means algorithm in the QuB software suite was utilized for HMM analysis using a two-state model to idealize the data. This data set contained in total 150 smFRET traces. All FRET values are rounded to two decimal digits.

\subsubsection*{BHG-HMM construction for \texttt{FRETtranslator}.}

To obtain the hidden states, we first generated the BHG of the \textit{Bsu} preQ\textsubscript{1}-riboswitch with methods provided in \cite{RN30} using the parameter -k, indicating that 
pseudoknotted structures are taken into consideration. Only the LMs with negative energies in the resulting BHG were selected since they are thermodynamically favorable. This 
left just four candidate structures. We emphasize that each candidate structure as an LM represents a basin (ensemble) of structures in the RNA folding landscape. 
The initial probabilities of the four candidate structures were computed according to the Boltzmann distribution as previously described. Due to the lack of experimental data to 
determine the BHG time-scalar $c_0$, we computed in total four different sets of transition probabilities with $c_0\in \{10,10^2,10^3,10^4 \}$ and $t=0.1$ seconds according to 
Eqn.~\ref{E:tranprob}. Note that the transition probabilities stay the same as the initial probabilities when $c_0>10^4$ since the system reaches thermodynamic equilibrium. 

The emission probability for each candidate structure is estimated based on a set of $\geq 350,000$ \textit{in silico} generated 3D structures from \texttt{Ernwin} in order to make sure the 
sampling is sufficient. 

Without considering the added noise, a total of 8 BHG-HMMs were generated from different parameter combinations for further data analysis with 
\texttt{FRETtranslator}.

\section*{Results}
We applied our method to a 36-nt \textit{Bsu} preQ\textsubscript{1} riboswitch. This RNA is known to preferentially adopt a pseudoknot structure  upon binding of a preQ\textsubscript{1} ligand. Previous 
smFRET studies \cite{RN19} have revealed a relative shift in FRET states from a partially folded conformation with a moderate FRET value of $\sim 0.6$ in the absence of the ligand to a fully folded 
conformation with a high FRET value of $\sim 0.9$ in the presence of ligand. Through biochemical and computational analysis, these FRET conformations were shown to correspond to hairpin and 
pseudoknotted structures, respectively \cite{RN19}.  

\begin{figure}[!h]
\centering
  \includegraphics[width=\textwidth]{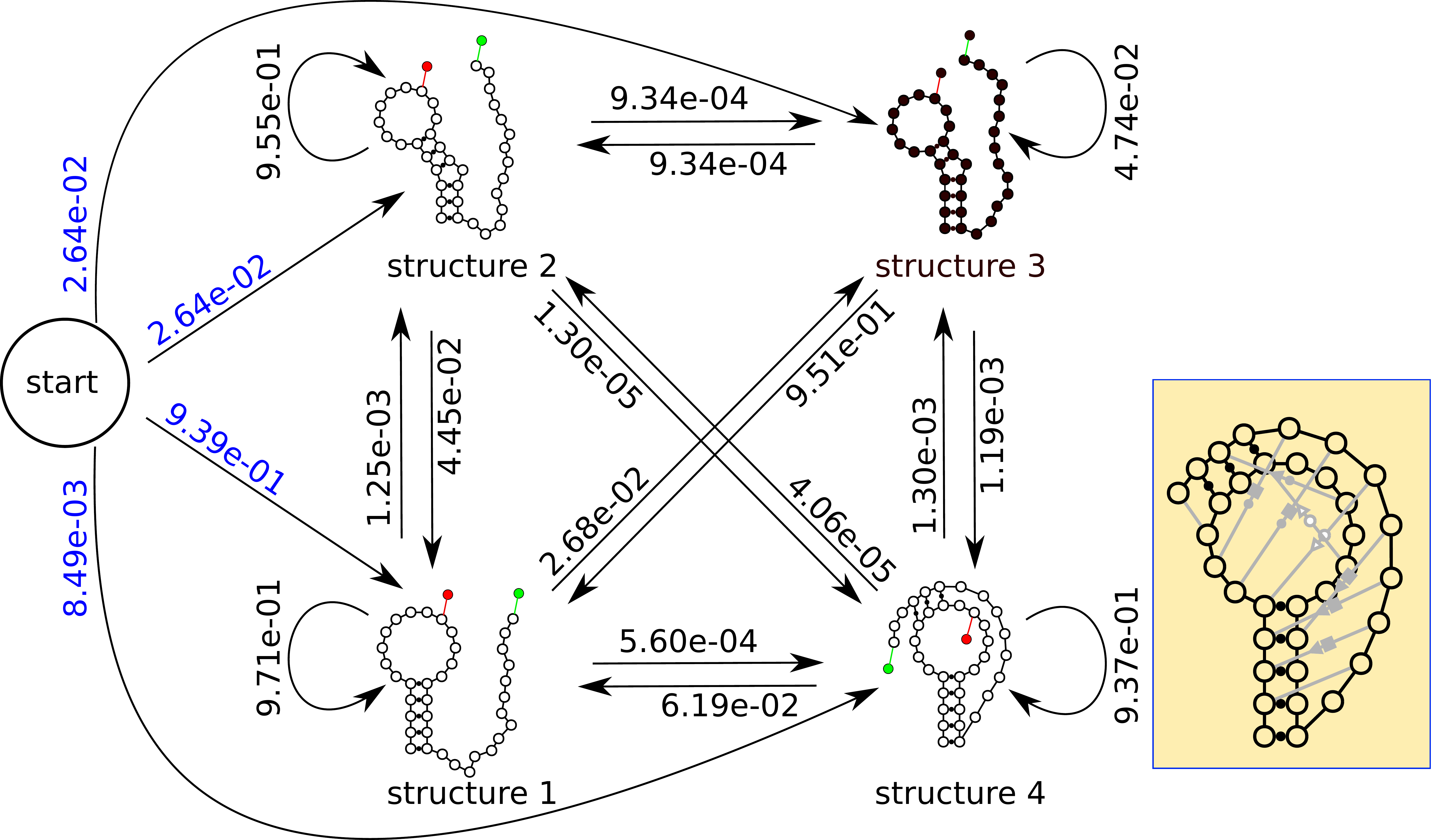}
\caption{{\bf Hidden states, initial probabilities and transition probabilities of a BHG-HMM constructed for the preQ\textsubscript{1} riboswitch. } 
Red and green markers indicate the locations of donors and acceptors, respectively. Arrows indicate possible transitions. The numbers in blue and black are the initial and transition 
probabilities, respectively. The transition probabilities are calculated with a BHG time-scalar $c_0=10$. The previously determined native structure in the ligand-bound state \cite{RN19} is 
indicated with a yellow background for comparison. Non-canonical base pairs in the native structure, which are not considered in the BHG model, are colored in grey. All diagrams of 
RNA secondary structures are plotted with RNAfdl \cite{RN43}. }
\label{F:preQ1hmm1}
\end{figure}

\begin{figure}[!h]
\centering
  \includegraphics[width=\textwidth]{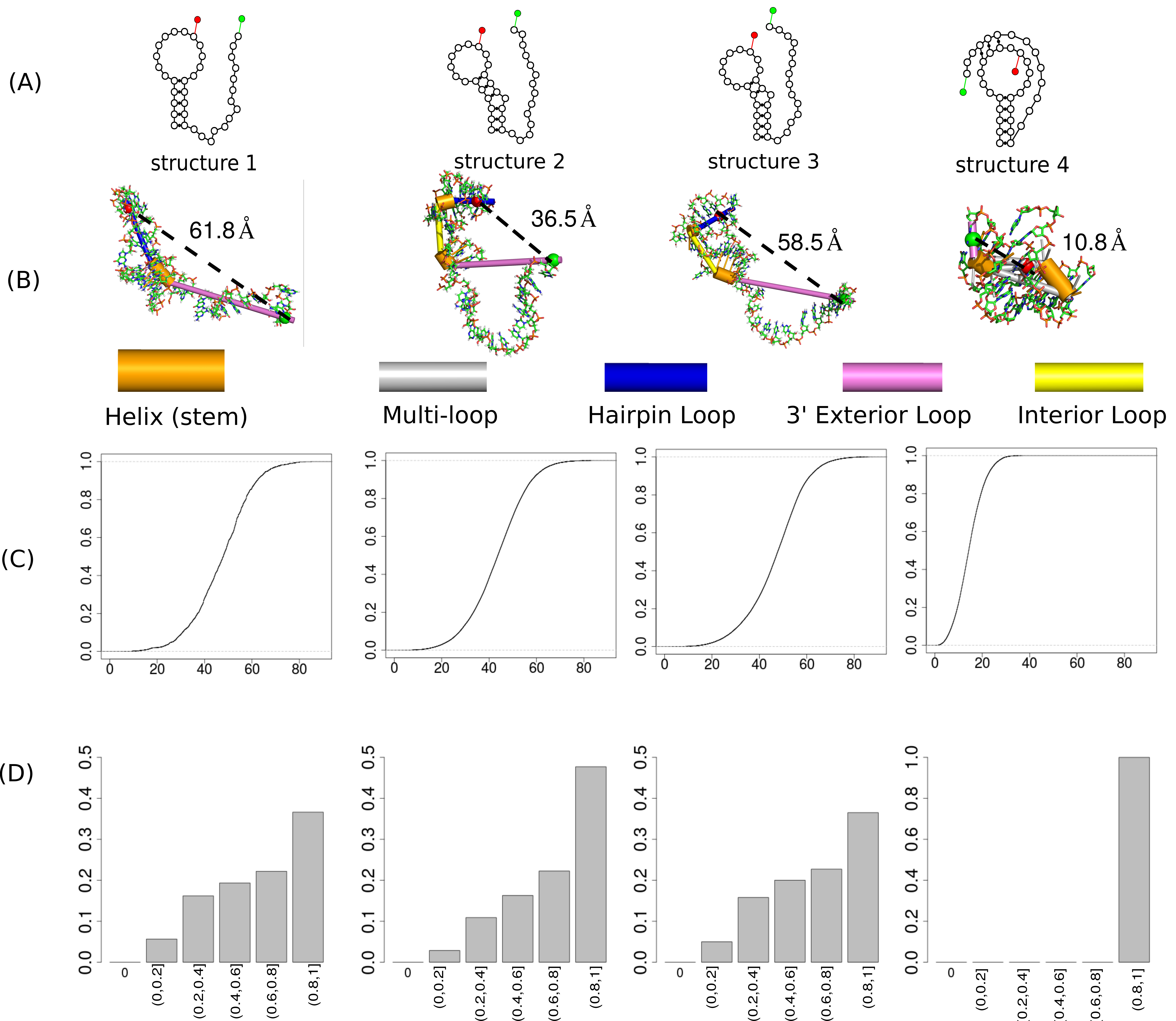}
\caption{{\bf Emission probabilities of a BHG-HMM constructed for the preQ\textsubscript{1} riboswitch.  } 
(A) Four candidate structures, where red and green markers indicate the nucleotide locations of the donor (C13) and accepter fluorophores (G36), respectively. (B) Examples of in 
silico generated 3D structures with \texttt{Ernwin} are displayed below their corresponding secondary structures. The structural elements in use are indicated below these structures. 
(C) The empirical cumulative distribution function of the \textit{in silico} distances for each candidate structure. (D) The emission probabilities are converted from \textit{in silico} distances 
estimated with \texttt{Ernwin}. The nonzero FRET values are grouped into 5 bins and an extra bin for value 0. Each bar indicates the emission probability of a bin, which is essentially an 
interval of FRET-values as labeled in the figure.}
\label{F:preQ1emission}
\end{figure}

As shown in Figs.~\ref{F:preQ1hmm1} and ~\ref{F:preQ1emission}, the computed BHG-HMM of this RNA contains four candidate structures. These structures are indexed according 
to their free energies in ascending order. Given that we only consider canonical base pairs, structure 1 matches the low-FRET hairpin structure whereas the base pairs of structure 4 are consistent with the 
previously determined pseudoknotted high-FRET structure \cite{RN19}.  The initial and transition probabilities between these structures are shown in Fig. 3, wherein the transition probabilities are 
calculated when the BHG time-scalar $c_0=10$. As shown in Fig.~\ref{F:preQ1emission}, the emission probabilities of candidate structures are estimated based on a large set of 
structures sampled with \texttt{Ernwin}. The emission probabilities of structure 1, 2 and 3 share a similar pattern whereas the emission probabilities of structure 4 are dominated by high FRET values.

\texttt{FRETtranslator} is applied to decode all 150 smFRET traces independently according to all 8 BHG-HMMs initially derived. In this step, no noise is added for estimating the emission 
probabilities. In general, as shown in Table~\ref{T:resultpreQ1nonoise}, when the BHG time-scalar $c_0=10, 10^2,10^3$, structural transitions between structures 1 and 4 are 
predicted to dominate. Such transitions occupy $\geq 96.5\%$ of the overall predicted transitions. Transitions between structures 1 and 2 are observed as well, but relatively rarely. 
When $c_0$ increases to $10^4$, no structural transition is observed, as for all time stamps among all traces the RNA is predicted to stay in structure 1. This is because the transition probability from structure 
4 to itself exhibits a sharp drop from $\geq 92\%$ to$ \sim 1\%$. In the meanwhile, the transition probability from structure 4 to structure 1 increases from $\leq 8\%$ to $94\%$. Thus, when $c_0=10^4$, 
instead of staying in structure 4, molecule prefers to quickly fold into its most stable structure, i.e., structure 1. That is, the molecule tends to stay in structure 1 rather than to transit to another structure.

\begin{table}[!ht]
\begin{adjustwidth}{-2.25in}{0in} 
\centering
\caption{\textbf{Numbers of structural transitions predicted for the preQ\textsubscript{1} riboswitch with different BHG-HMMs without added noise.}
Different BHG-HMMs are defined based on BHG time-scalar $c_0$  (first column) and the number of bins b (second column).  Values listed in column 3, 4, 5 and 6 are the percentages of the overall predicted transitions which are given in column 7. The overall ranking of the combinations of parameters based on their Viterbi scores are shown in the last column. Two optimal BHG-HMMs are highlighted in blue.}
\begin{tabular}{|cc+cccccc|}
\hline
$log(c_0)$	& $b$	& 1$\rightarrow$4 (\%)	& 4$\rightarrow$1 (\%)	&
1$\rightarrow$ 2 (\%) 	& 2$\rightarrow$1 (\%)	& Total number of
transitions	& Ranking\\ 
\thickhline
1	& 5	& 48.4	& 48.8	& 1.6	& 1.3	& 2608	& 2\\\hline
1	&10	& 48.5	& 49.1	& 1.2	& 0.1	& 2645	& 5\\\hline
\rowcolor{LightCyan}
2	& 5	& 48.9	& 49.2	& 1.0	& 1.0	& 2980	& 1\\\hline
\rowcolor{LightCyan}
2	&10	& 48.0	& 48.5	& 1.8 	& 1.7 	& 2947	& 4\\\hline
3	& 5	& 49.3	& 49.6	& 0.6 	& 0.5	& 3604	& 3\\\hline
3	&10 	& 48.1	& 48.6	& 1.7	& 1.7	& 2963	& 6\\\hline
\end{tabular}
\begin{flushleft} 
\end{flushleft}
\label{T:resultpreQ1nonoise}
\end{adjustwidth}
\end{table}

\begin{figure}[!h]
\centering
  \includegraphics[width=0.8\textwidth]{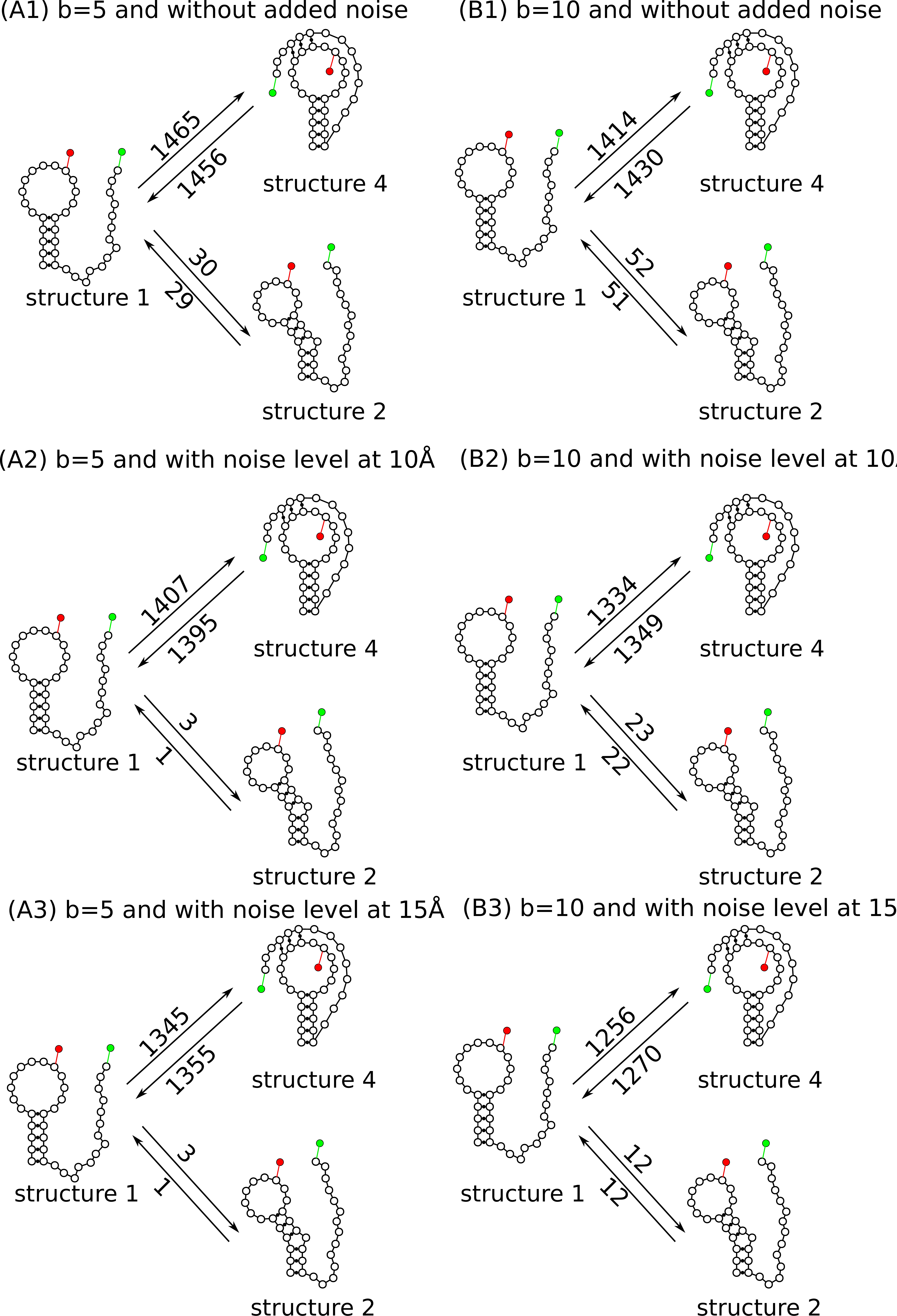}
\caption{
{\bf Structural transition predictions for the preQ\textsubscript{1} riboswitch using the optimal BHG-HMMs from the first round and their noise-added variants. } 
(Left) Optimal BHG-HMM with $b = 5$ (A1) and two noise-added variants:
using noise level $\Delta=10$ \AA (A2) and $\Delta=15$ \AA (A3). (Right)
Optimal BHG-HMM with $b=10$ (B1) and two noise-added variants: using
noise level $\Delta=10$ \AA (B2) and $\Delta=15$ \AA (B3). Structural transitions are dominated by transitions between structure 1 and structure 4 comparing to transitions between structure 1 and structure 2. 
The added noise reduces the number of the transitions between structure 1 and structure 2.
}
\label{F:preQ1freqtran}
\end{figure}

According to their Viterbi scores, two optimal BHG-HMMs (highlighted in blue in Table \ref{T:resultpreQ1nonoise}) are selected from the groups of BHG-HMMs based on bin sizes 
$b=5$ or $b=10$. For each optimal model, two levels of noise $\Delta=10$\AA and 15\AA are added to perturb the emission probabilities for testing their robustness in a second 
round of predictions. Fig.~\ref{F:preQ1freqtran} summarizes the structural transitions predicted with FRETtranslater in this second round, using the optimal first-round BHG-HMMs 
and their noise-added variants.

We observed that adding noise has two effects on the results. First, it reduces the total number of transitions predicted with \texttt{FRETtranslator} ($\leq 15\%$). In fact, the larger the 
value of $\Delta$ is, the fewer transitions are observed. This is mainly because added noise changes the emission probability distributions of structure 4 for the bins containing high 
FRET values. When no noise is added, the emission probability of structure 4 is nonzero only for the bins with FRET values in the range of $(0.8,1]$ when $b=5$ and $(0.9,1]$ 
when $b=10$. Secondly, noise reduces the number of transitions predicted between structures 1 and 2. Further, it substantially (by $\geq 60\%$) reduces the total number of occurrences 
of structure 2 as a predicted structure. This is mainly because noise
decreases the emission probabilities of structure 2 for FRET values
lying in the range of $(0.4,0.9]$. 

Given that noise naturally causes differences, we further sought to identify predictions that are noise-robust. In particular, when the number of bins is fixed, we only consider the 
predictions that are reported by all three cases: the optimal BHG-HMM and its two noise-added variants at noise levels 10 and 15 \AA. We summarize the noise-robust structural 
transitions in Table~\ref{T:preQ1summary}. In general, transitions between structures 1 and 4 are dominant ($\geq 99.0\%$). When b = 5 and b = 10 only one and 22 transitions, 
respectively, are 
predicted between structures 1 and 2.

\begin{table}[!ht]
\begin{adjustwidth}{-2.25in}{0in} 
\centering
\caption{
{\bf Numbers of noise-robust structural transitions predicted for the preQ\textsubscript{1} riboswitch.}
Values listed in column 2, 3, 4 and 5 are the percentages of the overall robust predicted transitions which are given in column 6. These values are derived using noise-perturbed variants
of the two optimal BHG-HMMs highlighted in Table~\ref{T:resultpreQ1nonoise}.}
\begin{tabular}{|c+ccccc|}
\hline
$b$	& 1$\rightarrow$4 (\%)	& 4$\rightarrow$1 (\%)	& 1$\rightarrow$ 2 (\%) 	& 2$\rightarrow$1 (\%)	& Total number of noise-robust transitions		\\ \thickhline
5	& 49.9	& 50.0	& 0.1		& 0.0		& 2630	\\\hline
10	& 49.3	& 49.8	& 0.5		& 0.4		& 2501  \\\hline
\end{tabular}
\begin{flushleft} 
\end{flushleft}
\label{T:preQ1summary}
\end{adjustwidth}
\end{table}

\begin{figure}[!h]
\centering
  \includegraphics[width=\textwidth]{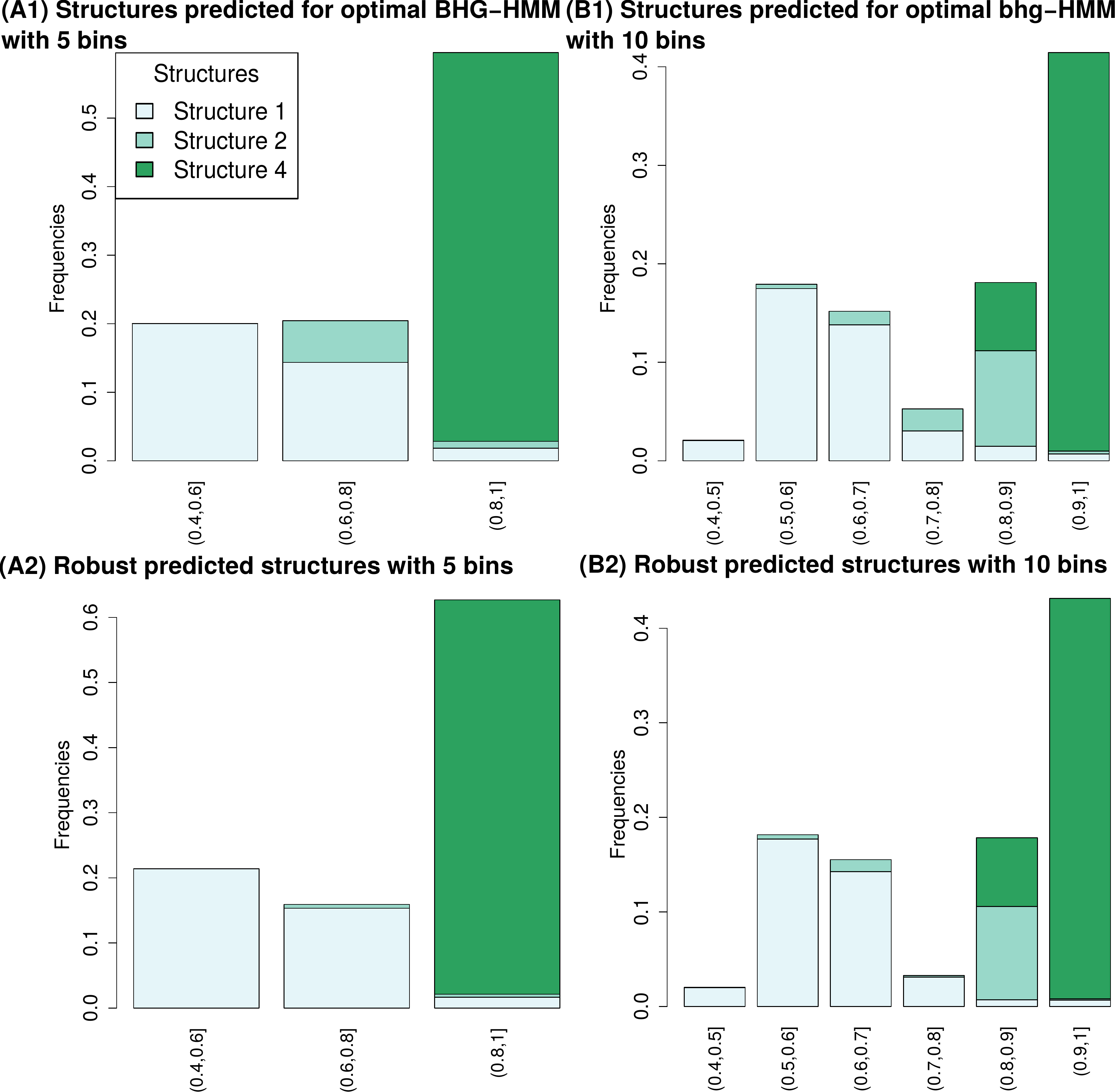}
\caption{
{\bf The compositions of predicted structures for all bin-determined FRET-value intervals. } 
When the numbers of bins are $b=5$ and $b=10$, the compositions of predicted structures for each bin-determined FRET-value interval using the optimal BHG-HMMs are shown in (A1) and (B1), respectively. 
Furthermore, analogous compositions of predicted structures, which are robust against added noise, are provided in (A2) and (B2), respectively.
The frequencies of FRET-value intervals are indicated in the y-axes. Colors are used to distinguish different predicted structures as indicated. Only the bin-determined intervals with non-zero frequencies are 
shown.}
\label{F:preQ1robust}
\end{figure}

Fig.~\ref{F:preQ1robust} shows the compositions of predicted structures for each bin determined FRET-value interval based on the optimal BHG-HMMs. In addition, we show the 
compositions that are noise-robust. First, FRET values collected from all 150 input FRET traces range from 0.46 to 0.98. The high FRET values appear more frequent than the low 
FRET values. When $b = 5$, using the optimal BHG-HMM, structures 1 and 4 completely occupy the lower FRET-values within range $(0.4,0.6]$ and higher FRET-values within 
$(0.8,1]$, respectively. Structure 2 is observed often (29.8\%) but still much less frequently than structure 1 (70.2\%) for the FRET-values within $(0.6,0.8]$. This is mainly because 
the emission probability for structure 2 corresponding to FRET-value interval $(0.6,0.8]$ is the maximum among all candidate structures. Thus, for an input trace with FRET 
values lying only within $(0.6,0.8]$, \texttt{FRETtranslator} has a tendency to assign structure 2 to the entire trace. This is why not many predicted structural transitions involve structure 2 
while a largely static structure 2 still occupies a portion in the overall composition. Only a small fraction (8.8\%) of these structure 2 predictions is noise-robust since the added noise 
weakens the privilege of structure 2 regarding emission probabilities.

When $b=10$, using the optimal BHG-HMM, structures 1 and 4 dominantly occupy the lower FRET-values within range $(0.4,0.7]$ and higher FRET-values within $(0.9,1]$, respectively. Structures 1 and 2 
are predicted with similar frequencies for the FRET values within $(0.7,0.8]$, with structure 2 being of a slightly lower frequency of $\sim 42.2\%$.  Again only a small fraction of these structure 2 predictions is 
noise-robust, only $\sim 6.9\%$. All three structures are observed for FRET values within $(0.8,0.9]$ while structures 2 and 4 are predicted much more often than structure 1. Furthermore, structure 2 is most 
frequently observed ($\sim 53.6\%$) for this FRET-value interval, and the prediction regarding this FRET-interval is relatively noise-robust.

In summary, using a robust two-step analysis, our predictions with \texttt{FRETtranslator} show the following: (1) structural transitions are most likely observed between structures 1 and 4, while transitions between 
structure 1 and structure 2 are rarely observed; (2) Independent of the choices of the number of bins, structures 1 and 4 dominate the FRET values within $(0.4,0.8]$ and $(0.9,1]$, respectively; (3) For the 
FRET values within $(0.8,0.9]$, when $b = 5$, \texttt{FRETtranslator} suggests structure 4 is most likely, while both structures 2 and 4 are suggested when b = 10. Given that the prediction of structure 2 is sensitive 
to the chosen number of bins, we are more confident with the structure 4 predictions. Thus, the results of our method are consistent with a previous analysis \cite{RN19}, wherein structure 1 as the hairpin-like 
structure was assigned to all low-FRET values and structure 4 as the pseudoknotted structure was assigned to all high-FRET values.

\section*{Discussion}
\texttt{FRETtranslator} is a Viterbi-decoding approach for predicting RNA structural pathways based on smFRET traces, the analysis of RNA secondary structure landscapes, 
and 3D structure prediction. \ignore{Additional information provided by RNA footprinting can also be included into this framework.} \texttt{FRETtranslator} only requires smFRET traces and the 
RNA sequence as input. To our knowledge, this is the first method that is capable of directly translating an input smFRET trace into a folding trajectory of secondary structures. We 
apply our method to the preQ\textsubscript{1} riboswitch example. The results are consistent with previous analysis provided in \cite{RN19}, wherein the hairpin-like structure and the 
pseudoknotted structure are assigned to all low-FRET values and all high-FRET values, respectively.

The performance of \texttt{FRETtranslator} depends on the accuracies of its three elementary parts: RNA folding landscape analysis, RNA 3D-structure sampling and the quality of the 
smFRET traces. 

The RNA folding landscape defines the candidate secondary structures that we can assign to time points in the smFRET traces. In this study, we use basin hopping graph (BHG) to 
coarse grain the folding landscape. Vertices in BHG correspond to the
``basins'' in the landscape and are represented by the LM 
structure at the bottom of the basin. While it would be ideal to use 
the entire set of LMs, this is computationally infeasible
given the number LMs grows as the square root of the number of structures
\cite{RN62}. To overcome this limitation, 
in this study we consider only the low-energy part of the landscape by restricting the energy of structures within certain threshold. In future work, we might avoid selecting LMs 
sharing very similar structural elements. Existing methods, for instance RNAshapes \cite{RN61} and RNAgraphdist \cite{RN47} can be promising.

In our method, the 3D sampling software in choice determines the quality of the estimated emission probability and thus also the performance of the \texttt{FRETtranslator}.  \texttt{Ernwin} \cite{RN29} is to our best knowledge the only 
coarse-grained fragment-assembly approach has ability to generate ensembles of structures competitive with the predictions of more sophisticated all-atom models such as FARNA \cite{RN48}.  Typical all-atom approaches to 
the prediction of RNA 3D structure yield modest accuracy for smaller molecules but suffer from extremely low accuracy for any structure beyond $\sim$30 nt in length. Other recently developed coarse-grained 
approaches, e.g. SimRNA \cite{RN49} are also promising however only one optimal structure is provided. The prediction accuracy can likely be improved if more experimental data regarding structural element 
can be included: as the existence of pseudoknots as the case for the preQ\textsubscript{1} riboswitch. \ignore{or chemical footprinting data \cite{RN50}}

\ignore{Footprinting data are used in this study for validating \texttt{FRETtranslator}. More precisely, footprinting activities provide a criterion for us to evaluate the performance of \texttt{FRETtranslator} under different BHG-
HMMs. This is because for the second example Ubc4 pre-mRNA, there is no available structural information for us to validate our results.} 

Given the recent development of tools for RNA secondary structure 
prediction to integrate chemical probing data such as footprinting reactivities \cite{RN50, RN51, RN52, RN53}, the current RNA secondary structure sampling during the BHG computation could be extended to 
take footprinting data into account. This could potentially improve the performance of \texttt{FRETtranslator} predictions. However, incorporating chemical probing data is not a trivial task. For example, several 
different 
structures could each show little overlap with the chemical probing data, but match the reactivities as an ensemble of structures. The emerging MCM \cite{RN27} approaches, however, hold the promise to 
disentangle at least the most prevalent alternative conformations.

The performance of \texttt{FRETtranslator} furthermore depends on the design of the smFRET experiment itself. This includes the locations of donor and the acceptor fluorophores, the FRET signal and its pre-
processing. Inter-fluorophore distances should effectively reflect the structural changes. Also, the method used for idealizing FRET values can potentially have a large impact on the predictions. At least we 
suggest, the number of FRET states selected in idealizing raw FRET values should be the same as the number of bins used in selecting BHG-HMMs. In addition, the way that traces are split is also delicate. 
For instance, if the pre-processing of raw FRET efficiencies yields traces that only contain a single value it is less likely that our method performs well. After all, the strength of \texttt{FRETtranslator} is in analyzing 
transitions between structures. 

Finally, the Viterbi model, which \texttt{FRETtranslator} is based on, has its shortcoming as it focuses on a single path. Similar to alternative structures, there will also be alternative paths. In our future study, we hint 
at considering posterior decoding as a possible improvement of our method.

\section*{Supporting Information}

\paragraph*{S1 Text.}
\label{S1_Text}
{\bf RNA folding landscapes and basin hopping graphs.} 

\paragraph*{S2 Text.}
\label{S2_Text}
{\bf Measuring the \textit{in silico} inter-fluorophore distance with \texttt{Ernwin}.} 

\section*{Acknowledgments}

We thank Dr. Krishna Suddala for providing smFRET data of the preQ\textsubscript{1} riboswitch. Also many thanks to Dr. Marcel Kucharik for computing basin hopping graphs for the preQ\textsubscript{1} riboswitch\ignore{ and the Ubc4 pre-mRNA}.

\section*{Author Contributions}

PFS, ILH, JG, NGW and JQ conceived and designed the experiments. MLK performed the smFRET experiments. NH, PK and JQ performed computational analysis. NH, MLK, PK, NGW and JQ analyzed the data. All coauthors contributed reagents/materials/analysis tools and wrote the paper together.

\section*{Funding Information}
Austrian Science Fund (FWF): [M1618-N28] to JQ. Deutsche Forschungs
Gemeinschaft  STA 850/15-1 to PFS. The Lundbeck Foundation, Innovation Fund
Denmark to JG, US National Institutes of Health grant R01GM098023 to NGW. 

\nolinenumbers

%
%
%
%
%
%
%

\bibliographystyle{plos2015}
\bibliography{FRET}

\end{document}